\documentclass[conference]{IEEEtran}
\IEEEoverridecommandlockouts
\usepackage{cite}
\usepackage{amsmath,amssymb,amsfonts}
\usepackage{algorithmic}
\usepackage{graphicx}
\usepackage{textcomp}
\usepackage{xcolor}
\usepackage{comment}
\usepackage{gensymb}
\usepackage{microtype}
\usepackage[free-standing-units,per-mode=repeated-symbol,mode=text,detect-weight=true,detect-family=true]{siunitx}
\usepackage{booktabs}
\usepackage{multirow}

\usepackage[normalem]{ulem}

\usepackage{pgfplots}
\usepgfplotslibrary{colorbrewer}
\pgfplotsset{width=\columnwidth,compat=1.7}
\usetikzlibrary{patterns}

\definecolor{celestialblue}{rgb}{0.29, 0.59, 0.82}
\definecolor{celadon}{rgb}{0.67, 0.88, 0.69}
\definecolor{ceil}{rgb}{0.57, 0.63, 0.81}
\definecolor{carrotorange}{rgb}{0.93, 0.57, 0.13}
\definecolor{carolinablue}{rgb}{0.6, 0.73, 0.89}
\definecolor{cerise}{rgb}{0.87, 0.19, 0.39}
\definecolor{darkspringgreen}{rgb}{0.09, 0.45, 0.27}
\definecolor{deepcarminepink}{rgb}{0.94, 0.19, 0.22}

\definecolor{color1}{HTML}{c7f9cc}
\definecolor{color2}{HTML}{80ed99}
\definecolor{color3}{HTML}{57cc99}
\definecolor{color4}{HTML}{38a3a5}
\definecolor{color5}{HTML}{22577a}

\def\BibTeX{{\rm B\kern-.05em{\sc i\kern-.025em b}\kern-.08em
    T\kern-.1667em\lower.7ex\hbox{E}\kern-.125emX}}
    
\DeclareSIUnit\bits{bits}
\DeclareSIUnit\gate{GE}

\newcommand{\aw}{\ensuremath{\text{AW}}}
\newcommand{\dw}{\ensuremath{\text{DW}}}
\newcommand{\iw}{\ensuremath{\text{IW}}}
\newcommand{\mot}{\ensuremath{\text{MOT}}}
\newcommand{\qtyadj}[2]{\num{#1}-\unit{#2}}

\newcommand{\mv}[1]{{\color{cyan}#1}}

\begin{document}
\bstctlcite{IEEEexample:BSTcontrol}

\title{PATRONoC: Parallel AXI Transport Reducing Overhead for Networks-on-Chip targeting Multi-Accelerator DNN Platforms at the Edge}

\author{
        \IEEEauthorblockN{Vikram~Jain$^\dagger$, Matheus~Cavalcante$^\star$, Nazareno~Bruschi$^\oplus$, Michael~Rogenmoser$^\star$, Thomas~Benz$^\star$, \\
        Andreas~Kurth$^\star$, Davide~Rossi$^\oplus$, Luca~Benini$^\star$$^\oplus$~and~Marian~Verhelst$^\dagger$ \\
        $^\dagger$ESAT-MICAS, KU Leuven, Belgium, $^\star$IIS, ETH Zurich, Switzerland, $^\oplus$University of Bologna, Italy\\ Email: vikram.jain@kuleuven.be}
        \vspace{-1.5em}
}

\maketitle

\begin{abstract}
Emerging deep neural network (DNN) applications require high-performance multi-core hardware acceleration with large data bursts. Classical network-on-chips (NoCs) use serial packet-based protocols suffering from significant protocol translation overheads towards the endpoints. This paper proposes PATRONoC, an open-source fully AXI-compliant NoC fabric to better address the specific needs of multi-core DNN computing platforms. Evaluation of PATRONoC in a 2D-mesh topology shows \qty{34}{\percent} higher area efficiency compared to a state-of-the-art classical NoC at \qty{1}{\giga\hertz}. PATRONoC's throughput outperforms a baseline NoC by \num{2}-\num{8}$\times$ on uniform random traffic and provides a high aggregated throughput of up to \qty{350}{\gibi\byte\per\second} on synthetic and DNN workload traffic.  
\end{abstract}


\section{Introduction}
Deep neural networks (DNNs) have become one of the primary workloads in  computing platforms of data centers and edge devices in the internet of things (IoT). Given the high proliferation of DNN workloads, research into designing and developing high-performance specialized hardware accelerators for DNN has gained much interest, as evidenced by the several DNN accelerators presented in the past decade~\cite{survey}. 
In the quest to support the ever-growing requirements of DNN workloads, hardware architectures have evolved from small single-core implementations to homogeneous~\cite{tinyvers} and heterogeneous~\cite{diana,esp_chip,kraken} multi-core hardware implementations\footnote{In this paper, the terms core and accelerator are used interchangeably.}. The trend of going multi-core can bring performance gains. However, it also brings new challenges, such as resource partitioning, workload mapping, complex hardware implementations, memory hierarchy design, and data communication bottlenecks between cores.

Multi-CPU-based general-purpose computing traditionally uses networks-on-chip (NoCs) and their various optimizations for inter-CPU data communication. Many topologies exist to balance the scalability of CPU cores, throughput, latency, and area impact of the NoC. Moreover, NoC protocols are designed for packetization and serialization over fairly narrow channels between cores (e.g., 32 bits), which reduces the number of routing resources needed. However, this implies additional hardware at the network's edges for protocol translation and serialization/deserialization (SERDES) from standard channel-oriented protocols at the endpoints (e.g., AXI4 or AXI5) to the NoC protocol. Moreover, due to their serialized nature, these NoCs need a high clock frequency to meet the bandwidth requirements, thus needing clock domain crossing hardware. 


Such traditional narrow-channel NoCs work well for inter-CPU cache traffic. However, the traffic of DNN workloads is mostly deterministic, with large bursts of non-coherent data movements requiring high bandwidth interconnection to achieve high performance and low latency. 
To achieve high bandwidth, typical solutions either 1) use a narrow NoC and operate it at 2-8$\times$ the core frequency~\cite{kaist} or 2) build a wide NoC with multiple channels~\cite{dally}. The latter solution gains traction as advanced technology scaling enables the area-efficient integration of more and more on-chip interconnect resources~\cite{dally,symbiosis}. However, modern NoCs need more than just wide links to answer the needs of DNN workloads, as packet-based serial NoC protocols are inadequate for workloads that rely on burst-based traffic.  




This paper proposes a template for burst-based homogeneous AXI-compliant NoCs to address the requirements of emerging multi-core DNN platforms and to tackle the challenges of packet-based serial NoCs. This work builds upon the open-source elementary AXI building blocks of~\cite{andy_paper}, which focuses on crossbar-based topologies, towards a fully-configurable open source AXI-based NoC framework, PATRONoC. PATRONoC is subsequently used in a mesh topology and extensively benchmarked to demonstrate
the benefits of having AXI-based NoCs. As such, this work makes the following contributions:

\begin{itemize}
    \item We present an open-source parameterizable AXI-compliant NoC designed for providing high bandwidth links for multi-core DNN computing platforms. The NoC is available at https://github.com/pulp-platform/axi.
    \item We demonstrate that using an AXI protocol for the NoC creates a fully homogeneous network interface to avoid high cost of protocol translation and provides a standard plug-and-play support for ease of integration.
    \item We show that using the AXI protocol end-to-end, a multi-channel, wide NoC with burst support and high bandwidth between cores as well as to-and-from memory can be supported, thereby improving performance of DNN applications on multi-core platforms.
\end{itemize}

The rest of the paper is organized as follows. Section~\ref{sec:axinoc} discusses the architectural overview of the proposed NoC, followed by details of the NoC's physical implementation with GlobalFoundries' modern 22FDX technology in Section~\ref{sec:impl}. We evaluate our NoC with synthetic and real traffic patterns extracted from DNN workloads in Section~\ref{sec:eval}. Subsequently, we compare our work against other modern NoC solutions in  Section~\ref{sec:relw} before presenting our conclusions in Section~\ref{sec:conc}.

\section{Interconnect Architecture of PATRONoC}
\label{sec:axinoc}

\begin{figure}[!t]
\includegraphics[width=\columnwidth]{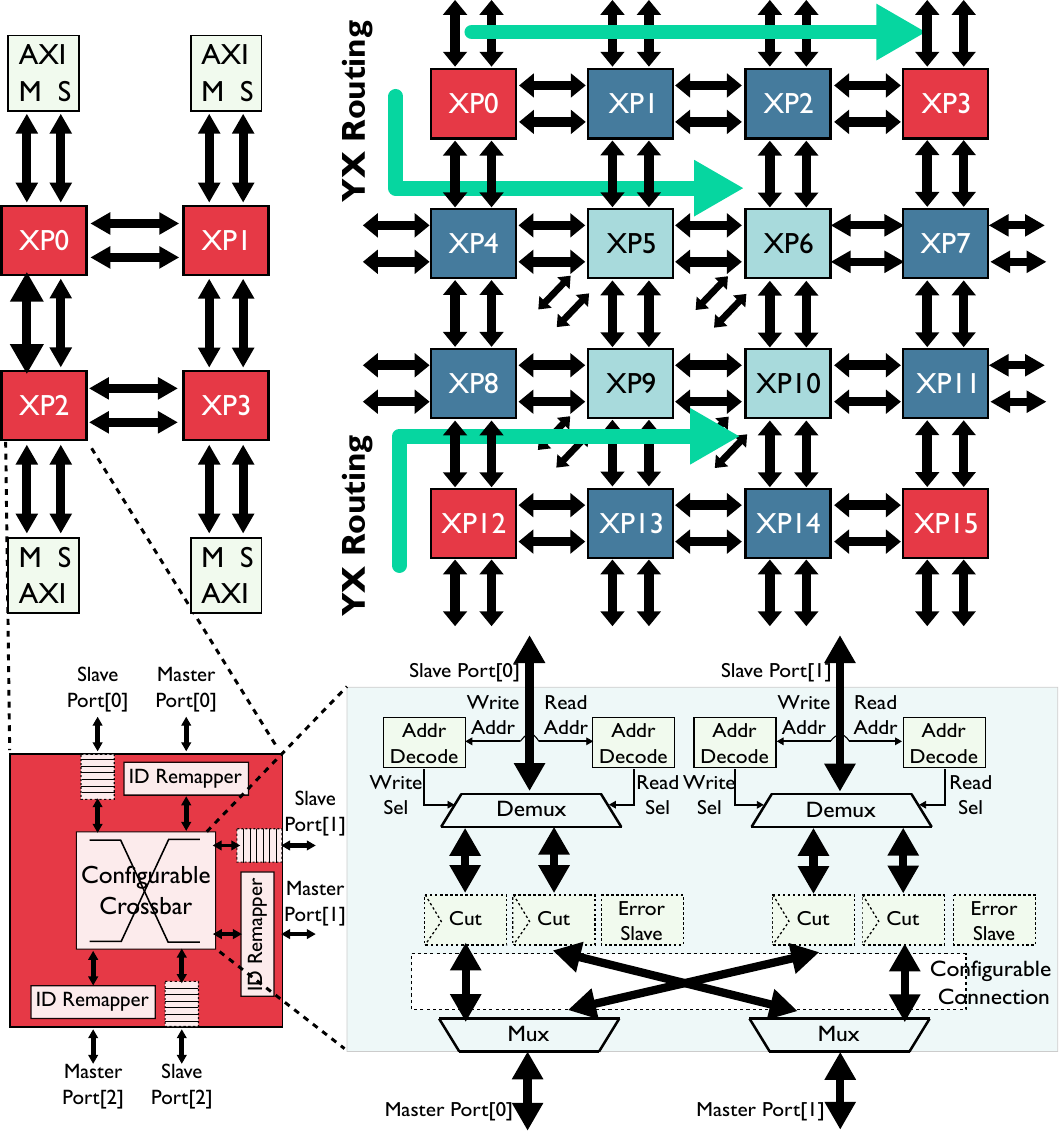}
\vspace{-2.0em}
\caption{PATRONoC instances as a \numproduct{2x2} mesh (left) and a \numproduct{4x4} mesh (right). The AXI masters and slaves are not shown in the \numproduct{4x4} mesh for ease of readability. Elementary blocks used for the NoC are also shown: XP (bottom-left) and XBAR (bottom-right). Red XP is 3-master and 3-slave, light blue XP is 4-master and 4-slave, and, dark blue XP is 5-master and 5-slave.
}
\vspace{-1.3em}
\label{fig_mesh}
\end{figure}

This section provides architectural and physical implementation details of PATRONoC for a mesh topology. 
NoCs are built with many elementary routing elements, each forwarding data from the ingress ports to the egress ports according to their topology-specific routing table. In this work, we extend the AXI crosspoint (XP) from~\cite{andy_paper}, shown in Fig.~\ref{fig_mesh} (bottom), allowing it to be used as PATRONoC's routing element. 
The XP consists of a configurable crossbar (XBAR) switch and ID remappers to ensure isomorphic XP ports. 
It is fully AXI-compliant and supports bursts, multiple outstanding transactions, and transaction ordering. We used the XP as the building block for a homogeneous, 2D mesh topology NoC with widely configurable dimensions, as shown in Fig.~\ref{fig_mesh}. Although this work uses the 2D mesh as a proof-of-concept, any regular topology, such as a torus, butterfly, or ring, can also be modularly built using our building blocks. We focused on the mesh due to its popularity in research and its remarkable simplicity, scalability, and efficiency~\cite{dally}. Fig.~\ref{fig_mesh} shows the two mesh topologies, \numproduct{2x2} (top-left) and \numproduct{4x4} (top-right), used to evaluate PATRONoC.

The meshes are built by instantiating the XPs in a 2D grid and connecting the NESW-bound links to neighboring XPs. AXI masters and slaves can be connected as NoC endpoints at each XP. A common AXI master is a core or a DNN accelerator, and AXI slaves can be memory or I/O tiles. Each XBAR is configured with a static routing table used for deterministic dimension-ordered routing in the mesh. Specifically, PATRONoC uses a source-based YX routing scheme, 
as shown with the green arrows in Fig.~\ref{fig_mesh}, to reduce the complexity of the route calculation step of the crosspoints. In this algorithm, a transaction is first passed forward in the same column until it reaches the same row as the destination XP and then passed forward in the same row until it reaches the destination XP. An automated script generates the address-based routing table for each XP which is used for routing the AXI transactions based on their destination address. 


\begin{scriptsize}
\begin{table}[!t]
  \centering
  \caption{Main parameters of the PATRONoC 2D mesh.}
  \label{table:params}
  \begin{tabular}{rl}
    \toprule
    \textbf{Parameter} & \textbf{Values}\\
    \midrule
    Mesh Dimension & N$\times$M \\
    Number of AXI Masters & 1 to N$\times$M (default)  \\
    Number of AXI Slaves & 1 to N$\times$M (default) \\
    Data Width & \qtyrange{8}{1024}{\bits} \\
    Address Width & Arch. dependent (32 or \qty{64}{\bits}) \\
    ID Width & \qty{1}{\bit}~to~\qty{16}{\bits} \\
    Max \#Outstanding Trans. & 1 to 128 \\
    XBAR Connectivity & Partial (default) or Fully connected \\ 
    Register Slice & Single channel or all channels (default) \\
    \bottomrule
  \end{tabular}
\vspace{-1.5em}
\end{table}
\end{scriptsize}

PATRONoC is highly parameterizable, taking advantage of the flexibility of the AXI protocol. The parameters that can be tuned at design time are shown in Table~\ref{table:params}. 
The number of AXI masters and slaves indicate the number of connected cores and memory/IO tiles in the design. Both ranges for possible number of AXI masters and slaves are valid for the N$\times$M 2D mesh and are topology-dependent. For example, in a concentrated mesh, multiple masters and slaves can connect to the same XP. Furthermore, the data width (\dw{}) can be tuned to meet the system's bandwidth requirements, while the address width (\aw{}) can be tuned to support a larger global address space. 

The AXI protocol identifies transactions with IDs used by the master endpoints to distinguish independent transactions. The number of unique IDs can be configured using the ID width (\iw{}) and increases with the number of masters. All transactions from the same master with the same ID must remain ordered, but there is no ordering requirement between transactions with different IDs. Multiple outstanding transactions enable the master to hide the memory latency. A higher max.\ number of outstanding transactions (\mot{}) improves performance, as all AXI building blocks can support multiple in-flight transactions, preventing bandwidth degradation when the NoC is saturated. 

The XBAR connectivity parameter configures the XP to either connect all slave ports to all master ports in the case of a fully-connected network or partially connect slaves and masters in the case of a mesh or other non-point-to-point topologies. The last parameter is the register slice (cut), shown in Fig.~\ref{fig_mesh}, that can be optionally inserted at design time on some or all AXI channels, improving the timing of the design at the cost of increased latency. The rest of the paper evaluates PATRONoC in \numproduct{2x2} and \numproduct{4x4} mesh topologies with multiple configurations based on the \dw{}, \aw{}, \iw{}, and \mot{} parameters.




\section{Implementation Results}
\label{sec:impl}

This section provides implementation results in terms of complexity and scalability of the NoC and its parameters. The implementation is done in GlobalFoundries' 22FDX technology node using a ten-layer metal stack. We used eight-track standard cells of SLVT/LVT type, characterized at worst-case scenario (SS/\qty{0.72}{\volt}/\qty{125}{\celsius}) for timing analysis. The designs from Section~\ref{sec:axinoc} are synthesized using Synopsys' Design Compiler 2022.03 in topographical mode, taking physical endpoint placement constraints into account. 
All designs achieve a clock frequency of \qty{1}{\giga\hertz} at the worst-case condition corner with a register slice on every AXI channel. 


\begin{figure}[!t]
\centering
\includegraphics[width=0.9\columnwidth]{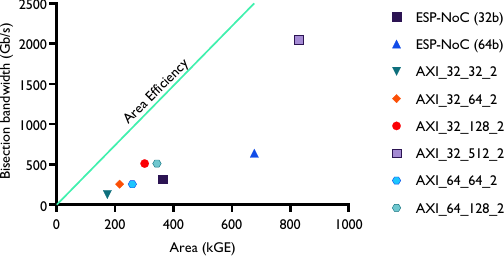}
\vspace{-1.0em}
\caption{Implementation results showing area versus bisection bandwidth of PATRONoC and ESP-NoC~\cite{esp} in \numproduct{2x2} mesh configurations. PATRONoC's configurations are represented as AXI\_AW\_DW\_IW.}
\vspace{-1.5em}
\label{fig_impl_2x2_mesh}
\end{figure}


The \numproduct{2x2} PATRONoC mesh, shown in Fig.~\ref{fig_mesh} (top-left), is first synthesized with different \aw{} and \dw{} parameters, keeping $\iw{}=\qty{2}{\bits}$, $\mot{}=1$, and other parameters at default values. 
Fig.~\ref{fig_impl_2x2_mesh} shows the area versus bisection bandwidth (\dw{}-dependent) of the mesh for the different configurations. As expected, the design area scales up with increasing \aw{} and \dw{}, taking up mere \qty{174}{\kilo\gate} for the smallest configuration of $\aw{}=\qty{32}{\bits}$ and $\dw{}=\qty{32}{\bits}$. The biggest design shown in Fig.~\ref{fig_impl_2x2_mesh}, with $\dw{}=\qty{512}{\bits}$, takes an on-chip area of \qty{830}{\kilo\gate}.

The benefit of having a homogeneous NoC is evident when the design is compared to classic NoC solutions. This work uses ESP-NoC~\cite{esp} as our baseline NoC. ESP-NoC is a state-of-the-art open-source packet-based NoC including six planes for coherent and non-coherent traffic for multi-core heterogeneous systems. Synthesis results showing the area of the \numproduct{2x2} ESP-NoC mesh in its \qtyadj{32}{\bit}- and \qtyadj{64}{\bit}-flit configurations are presented in Fig.~\ref{fig_impl_2x2_mesh}. Compared to PATRONoC's  configuration with $\aw{}=\qty{32}{\bits}$ and $\dw{}=\qty{64}{\bits}$, ESP-NoC takes up \qty{68}{\percent} more area to provide only \qty{25}{\percent} more throughput (five \qtyadj{32}{\bit} wide planes providing \qty{160}{\giga\bit\per\second}). The area overhead can be attributed to ESP-NoC's multiple planes with large protocol translation interfaces at each endpoint. The advantage of PATRONoC is much more evident in Fig.~\ref{fig_impl_2x2_mesh} when comparing its area efficiency (slope) with ESP-NoC. We define area efficiency as the bisection bandwidth normalized to the standard cell area, providing a measure of NoC performance at a given complexity. Fig.~\ref{fig_impl_2x2_mesh} shows that PATRONoC is closer to the Pareto front providing better area efficiency compared to the ESP-NoC in \qtyadj{32}{\bit} and \qtyadj{64}{\bit} configurations. 

\begin{figure}[!t]
\centering
\includegraphics[width=\columnwidth]{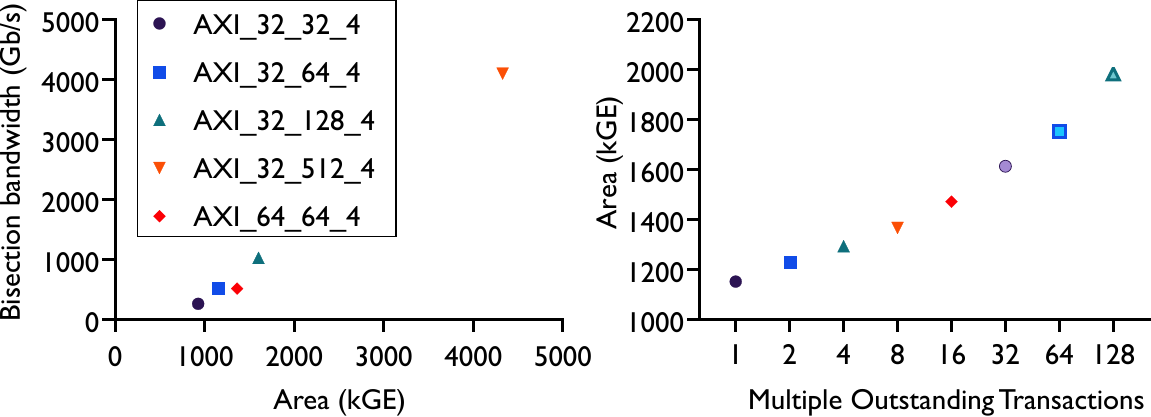}
\vspace{-1.5em}
\caption{Implementation results showing area vs. bisection bandwidth of PATRONoC in \numproduct{4x4} mesh configurations (left). Configurations are represented as AXI\_AW\_DW\_IW. Area vs. \mot{} tradeoff for $\dw{}=\qty{64}{\bits}$ (right).} 
\vspace{-1.5em}
\label{fig_impl_4x4_mesh}
\end{figure}

We implement the \numproduct{4x4} mesh shown in Fig.~\ref{fig_mesh} (top-right) to show the scalability of PATRONoC. For building the \numproduct{4x4} mesh, the \iw{} of the AXI blocks is increased to 4 to support 16 unique IDs required for 16 masters. The results of the area and bisection bandwidth of this mesh are summarized in Fig.~\ref{fig_impl_4x4_mesh} (left). As the mesh dimensions change, the area overhead of the NoC becomes approximately \qty{32}{\percent} compared to the \numproduct{2x2} mesh in similar \aw{} and \dw{} configurations, leading to a drop in area efficiency by \qty{25}{\percent}. 
Increasing the \mot{} improves the performance of the NoC at the cost of larger complexity in terms of area. Fig.~\ref{fig_impl_4x4_mesh} (right) shows the tradeoff between \mot{} and the area of the \numproduct{4x4} PATRONoC with $\dw{}=\qty{64}{\bits}$. While this work focuses more on performance and area aspects of the NoC, the power consumption at \qty{1}{\giga\hertz} for the \numproduct{4x4} PATRONoC is \qty{45}{\milli\watt} (for $\dw{}=\qty{32}{\bits}$) and \qty{171}{\milli\watt} (for $\dw{}=\qty{512}{\bits}$) on uniform random traffic. This accounts for less than \qty{10}{\percent} of the projected power consumption of a complete platform, assuming that a typical DNN accelerator connected to one NoC node uses \qtyrange{100}{200}{\milli\watt}.

\section{Performance Evaluation}
\label{sec:eval}

PATRONoC's performance is characterized in terms of throughput versus injected load through a cycle-accurate register-transfer level (RTL) simulation. 
This section evaluates the performance of the \numproduct{4x4} PATRONoC mesh in two configurations: 1) as a slim NoC with $\dw{}=\qty{32}{\bits}$ and 2) as a wide NoC with $\dw{}=\qty{512}{\bits}$, both with $\aw{}=\qty{32}{\bits}$, $\iw{}=\qty{4}{\bits}$, and $\mot{}=8$. Each master is a DMA engine, and the slaves are AXI-capable memories that cater to the DMA requests. The configurable and workload-specific maximum burst length is used by the RTL model of the DMA engine to create AXI-compliant bursts (adhering to address boundaries and max number of beats) for the NoC.
In our evaluation framework, the workload-specific burst length is randomized within a user-defined range to emulate a random burst length with a random source and destination address, while the bursts in the NoC are subject to AXI compliance. All analyses assume a clock frequency of \qty{1}{\giga\hertz} for the endpoints and the NoCs.



\subsection{Uniform Random Traffic}

\begin{figure}[!t]
\centering
\includegraphics[width=\columnwidth]{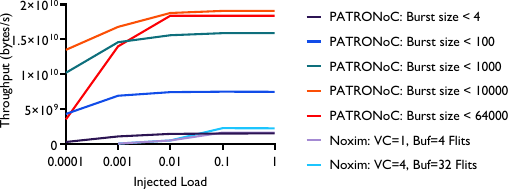}
\vspace{-1.0em}
\caption{Uniform Random Traffic with Poisson distribution using Noxim simulator for a \numproduct{4x4} 2D mesh and uniform random traffic on the slim PATRONoC with increasing DMA burst length.} 
\vspace{-1.5em}
\label{fig_noxim}
\end{figure}


The Noxim simulator~\cite{noxim} is used to set the baseline NoC performance, taking a \numproduct{4x4} mesh with the default XY routing, \qtyadj{32}{\bit} flits, and eight flits per packet 
to closely match the slim PATRONoC configurations. Fig.~\ref{fig_noxim} shows the non-exhaustive characterization of the Noxim NoC on this traffic in two configurations: 1) a standard single virtual channel with \num{4} flits per router buffer for a compact implementation, and 2) four virtual channels with \num{32} flits per router buffer for high performance. The saturation throughput of the Noxim NoCs are \qty{1.6}{\gibi\byte\per\second} and \qty{2.25}{\gibi\byte\per\second}, respectively. Increasing the number of virtual channels (VCs)~\cite{noc_book1} and flits per buffer improves the NoC's performance, but also increases router complexity.

Fig.~\ref{fig_noxim} also shows the NoC throughput for the uniform random traffic running on the \numproduct{4x4} slim PATRONoC mesh. It is clear that PATRONoC is beneficial for burst traffic. 
At small transfer lengths of less than \qty{4}{\byte}, 
similar to normal CPU traffic, PATRONoC performs equivalently to the Noxim NoC with \qty{1.5}{\gibi\byte\per\second} throughput. However, when using longer bursts, PATRONoC's performance improves and reaches up to \qty{19}{\gibi\byte\per\second} aggregated throughput at DMA burst lengths up to \qty{10}{\kibi\byte} and \qty{64}{\kibi\byte}. This provides an improvement of 8.4$\times$ over the saturation throughput achieved by the best Noxim NoC configuration (4 VCs, buffers 32-flit deep), showing that PATRONoC largely outperforms it by using bursts. 

\subsection{Synthetic Traffic}

\begin{figure}[!t]
\centering
\includegraphics[width=\columnwidth]{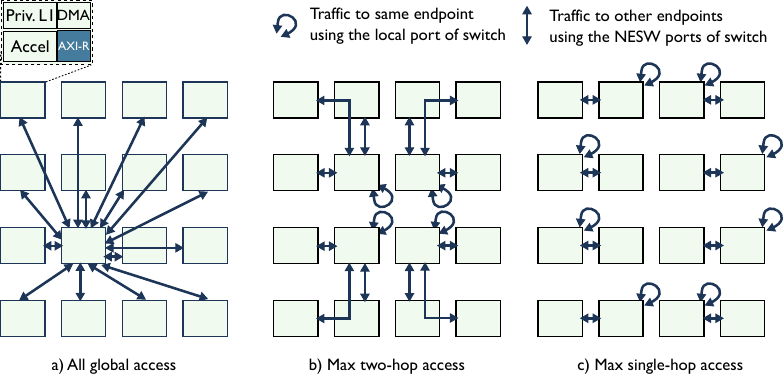}
\caption{Synthetic traffic patterns for the performance evaluation.}
\vspace{-1.5em}
\label{fig_use_cases}
\end{figure}

Fig.~\ref{fig_use_cases} shows the three synthetic traffic patterns considered: 1) all global access, 2) max two-hop access, and, 3) max single-hop access. 
We characterize the \numproduct{4x4} PATRONoC mesh in both slim and wide configurations with the synthetic patterns. 


\textit{a.) All global access}: In this traffic pattern, all the AXI master and DMA endpoints communicate with a single slave endpoint leading to predominately global accesses. Fig.~\ref{fig_use_cases}a) shows this traffic pattern on the \numproduct{4x4} mesh, where the endpoint $(2, 1)$ acts as the AXI slave. 
\textit{b.) Max two-hop access}: In this use case, the AXI slave accesses are distributed to four endpoints $(1, 1)$, $(1, 2)$, $(2, 1)$, and $(2, 2)$. This considers architectures that have a distributed shared L2/L1 memory, either uniform or non-uniform. The \num{16} AXI masters can communicate to any of the four endpoints, but in this case, the masters are restricted to only communicate to slaves which are a maximum of two hops away. 
\textit{c.) Max single-hop access}: In this traffic pattern, the AXI slaves are further distributed across eight endpoints along the edges except for the corners. The \num{16} masters are restricted to access only slaves which are at most one hop away. 
The last two cases are considered because in traffic from many DNN workloads, data scheduling can be done on nearby cores to prevent long latency and low-performance data communication.


\begin{figure}[!t]
\centering
\includegraphics[width=\columnwidth]{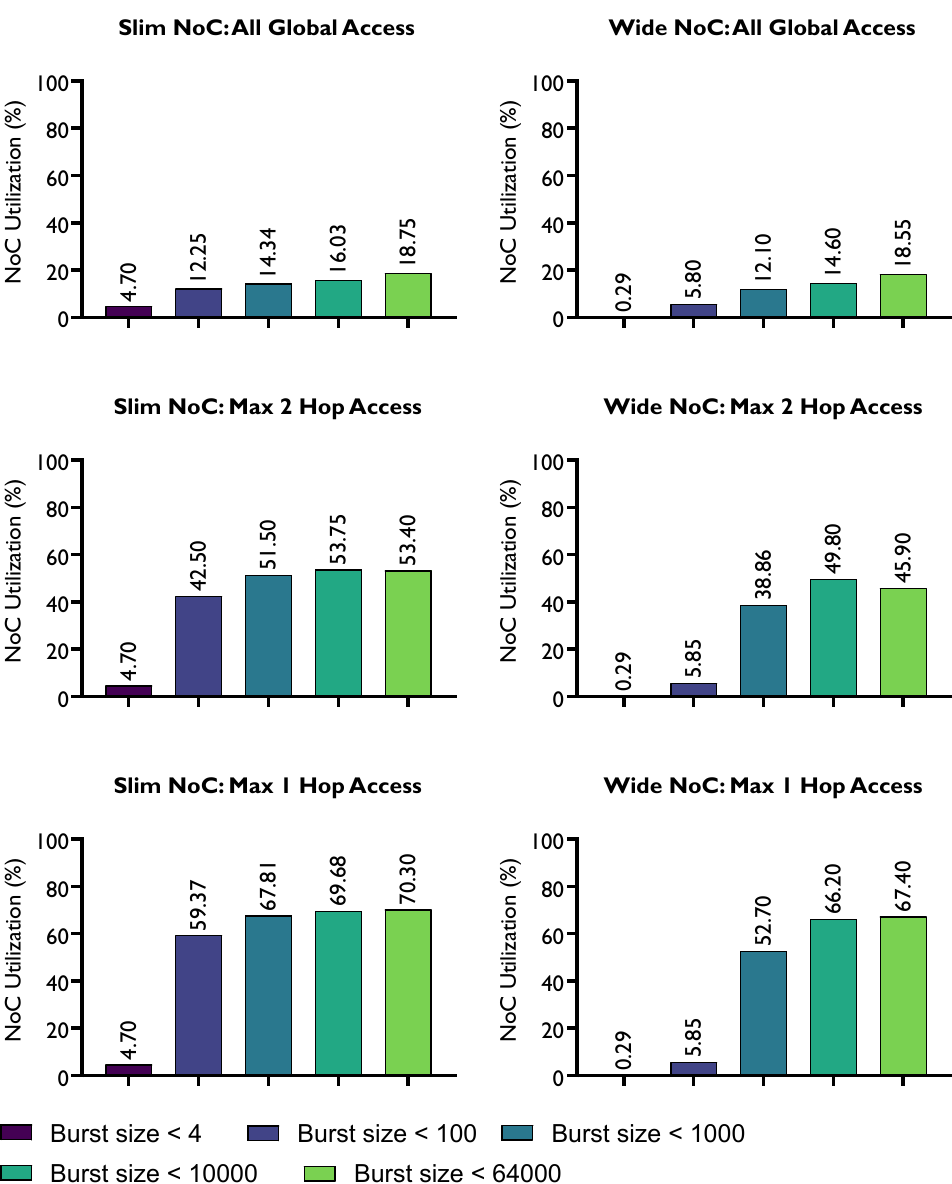}
\caption{NoC utilization at maximum injected load for the synthetic random traffic running on the slim and wide PATRONoC using all global access, max 2 hop, and max 1 hop traffic patterns with different DMA burst sizes.} 
\vspace{-0.5em}
\label{fig_synth_traf}
\end{figure}

The slim NoC can be used in architectures that are area-constrained but require more throughput than what most traditional NoCs can provide. Fig.~\ref{fig_synth_traf} (left) shows the NoC utilization, with respect to bisection bandwidth, of the slim NoC on the three synthetic patterns at different burst sizes. Starting with traffic pattern a.), the slim NoC provides a minimum of \qty{1.5}{\gibi\byte\per\second} of throughput with short bursts. This is approximately \qty{4.7}{\percent} NoC utilization considering the slim NoC has a \qty{32}{\gibi\byte\per\second} bisection bandwidth. The access pattern limits the traffic to a few links of the NoC and, thus, a low utilization is expected. The throughput improves considerably with increasing burst length and reaches a maximum of \qty{6}{\gibi\byte\per\second} for burst lengths up to \qty{64}{\kibi\byte}, providing a NoC utilization of around \qty{18.75}{\percent}. 
For pattern b.), the NoC performs similarly to pattern a.) for short burst lengths. However, the aggregated throughput improves considerably with larger bursts and saturates at \qty{17.2}{\gibi\byte\per\second} for burst lengths up to \qtylist{10;64}{\kibi\byte}. This leads to a higher NoC utilization of about \qty{53.75}{\percent}, showing that all mesh links can be utilized more efficiently. Similar to pattern b.), the pattern c.) under-performs at small bursts but the aggregated saturation throughput at larger bursts improves to \qty{22.5}{\gibi\byte\per\second} for bursts up to \qty{64}{\kibi\byte} with a NoC utilization of \qty{70.3}{\percent}. 

The wide NoC is geared towards high-bandwidth large-burst multi-core DNN-workload traffic. A significant performance gain can be achieved with such wide NoC, but being parameterizable means that also alternative \dw{}s between \qtylist{32;512}{\bits} can be considered by designers to find an optimal size for given system requirements. Fig.~\ref{fig_synth_traf} (right) shows the NoC utilization characteristic of the wide NoC on the synthetic access patterns with different burst sizes. 
For the traffic pattern a.), the wide NoC can only achieve a utilization of \qty{0.29}{\percent} at small bursts up to \qty{4}{\byte} large, providing a maximum throughput of \qty{1.5}{\gibi\byte\per\second} (bisection bandwidth of \qty{512}{\gibi\byte\per\second}). As seen with the slim NoC, this is an expected performance degradation with this access pattern. The degradation in NoC utilization is further exacerbated by the wide \dw{}s but short burst lengths. The throughput improves, however, with larger burst sizes of up to \qty{64}{\kibi\byte} and reaches saturation at \qty{95}{\gibi\byte\per\second} with \qty{18.55}{\percent} NoC utilization. Both patterns b.) and c.) result in low throughput and utilization with small burst sizes. The aggregated throughput improves at larger bursts with length up to \qtylist{10;64}{\kibi\byte} reaching \qty{255}{\gibi\byte\per\second} (\qty{49.8}{\percent} utilization) and \qty{345}{\gibi\byte\per\second} (\qty{67.4}{\percent} utilization) for the patterns b.) and c.), respectively.

\subsection{DNN Workload Traffic}

\begin{figure}[!t]
\centering
\includegraphics[width=\columnwidth]{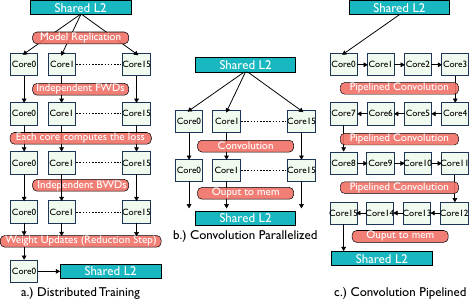}
\caption{Overview of the DNN workloads used for PATRONoC evaluation. FWD and BWD in (a) represents the forward and backward propagation workloads, respectively, used in DNN training.}
\vspace{-0.5em}
\label{fig_real_traffic}
\end{figure}

Synthetic traffic does not capture the full scope of access patterns in real multi-core hardware architectures running DNN workloads. In order to characterize the NoC in more realistic use cases, this section evaluates three emulated CNN-based workloads: a.)~distributed training, b.)~parallelized convolutions, and c.)~pipelined convolutions, shown in Fig.~\ref{fig_real_traffic}. We use GVSoC~\cite{bruschi2021gvsoc} to generate real traffic patterns for the RTL simulation. GVSoC is an open-source, highly configurable, and event-driven simulator for heterogeneous RISC-V-based SoC platforms used for full-system software development and performance evaluation. 

\textit{a.)~Distributed training}: 
For this workload, we replicate and deploy a ResNet-34 (\qty{90}{\percent} channel shrink factor) distributed training model for the ImageNet dataset on 16 cores. In terms of data communication, a mix of L2 to L1 (core), L1 (core) to L2, and L1 (core) to L1 (core) transfers are needed. \textit{b.)~Parallelized convolution}~\cite{workload}: This is a CNN-based inference workload in which the layers of the network and inputs are tiled and deployed on separate cores. This is a pure L2 to L1 (core) and L1 (core) to L2 memory traffic pattern and has no inter-core communication. \textit{c.) Pipelined convolution}~\cite{workload}: Depth-first or pipeline dataflow is used in many new DNN platforms to efficiently run CNN-based inference. In this scheme, layers are executed in parallel, in a pipelined way across the different cores to reduce the data traffic to higher memory levels~\cite{koen}.
This workload has mostly L1 (core) to L1 (core) traffic and only cores \num{0} and \num{15} do L1 (core) to/from L2 transfers.

\begin{figure}[!t]
\begin{tikzpicture}
\tikzstyle{every node}=[font=\small]
\centering
\begin{axis}[
    width=\columnwidth,
    height=4.5cm,
    ybar,
    enlarge x limits=0.6,
    ymin = 0,
    bar width=0.65cm,
    legend style={at={(0.45,-0.2)},
      anchor=north,legend columns=-1},
    ylabel={Throughput (\unit{\gibi\byte\per\second})},
    symbolic x coords={Slim NoC, Wide NoC},
    xtick=data,
    axis x line*=bottom,
    axis y line*=left,
    nodes near coords,
    nodes near coords align={vertical},
    legend image code/.code={
        \draw [#1] (0cm,-0.1cm) rectangle (0.2cm,0.25cm); },
    ]
\addplot [draw=gray,fill=color3] coordinates {(Slim NoC,5.176) (Wide NoC,83.1)};
\addplot [draw=gray,fill=color4] coordinates {(Slim NoC,4.266) (Wide NoC,68.5)};
\addplot [draw=gray,fill=color5] coordinates {(Slim NoC,19.173) (Wide NoC,310.7)};
\legend{Conv Train,Par Conv,Pipe Conv}
\end{axis}
\end{tikzpicture}
\caption{Throughput analysis for DNN workload traffic on the PATRONoC.}
\vspace{-0.5em}
\label{fig_real_slim}
\end{figure}
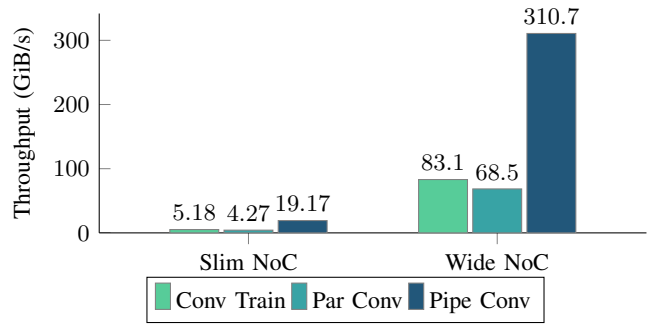

Fig.~\ref{fig_real_slim} shows the evaluation results of the \numproduct{4x4} slim and wide NoCs running the three DNN workloads. For the slim NoC, the parallelized convolution---which consists of mostly core to/from shared memory transfers---reaches a throughput of \qty{4.27}{\gibi\byte\per\second}. For the training workload, the throughput is better than the parallelized convolution workload as it involves a mix of core to/from shared memory and core-to-core transfers. On the pipelined convolution, which consists of predominantly core-to-core traffic, the NoC achieves a high \qty{19.17}{\gibi\byte\per\second} throughput. Similar trends are reported for the \numproduct{4x4} PATRONoC wide NoC shown in Fig.~\ref{fig_real_slim} (right), but at much higher throughput, with pipeline convolution reaching a peak throughput of \qty{310}{\gibi\byte\per\second}.

\begin{table}[!t]
\caption{Comparison of PATRONoC with state-of-the-art NoCs in SoCs}
\vspace{-2em}
\begin{center}
\begin{tabular}{rcccccc}
\toprule
&\multicolumn{5}{c}{\textbf{Metric}} \\
\cmidrule{2-6}
\textbf{Work} & \textbf{Open}& \textbf{Full} & \textbf{Burst-} & \textbf{Config-} & \textbf{NoC-BW} \\
& \textbf{Source}& \textbf{AXI} & \textbf{support} & \textbf{urable} & \textbf{(Gbps)$^{\mathrm{*}}$}\\
\midrule
SpiNNaker~\cite{spinnaker} & {\color{red}$\times$} & {\color{red}$\times$} & {\color{red}$\times$} & {\color{red}$\times$} & 5 (async) \\
Reza et al~\cite{cmesh} & {\color{red}$\times$} & {\color{red}$\times$} & {\color{red}$\times$} & {\color{red}$\times$} & 4000 \\
MCM~\cite{mcm} & {\color{red}$\times$} & {\color{red}$\times$} & {\color{red}$\times$} & {\color{red}$\times$} & 35 \\
MC-NoC~\cite{kaist} & {\color{red}$\times$} & {\color{red}$\times$} & {\color{red}$\times$} & {\color{red}$\times$} & 2368\\
NeuNoC~\cite{neunoc} & {\color{red}$\times$} & {\color{red}$\times$} & {\color{red}$\times$} & {\color{red}$\times$} & - \\
TETRIS~\cite{tetris} & {\color{red}$\times$} & {\color{red}$\times$} & {\color{red}$\times$} & {\color{red}$\times$} & - \\
PUMA~\cite{puma} & {\color{red}$\times$} & {\color{red}$\times$} & {\color{red}$\times$} & {\color{red}$\times$} & -\\
OpenSoC~\cite{opensoc} & {\color{darkspringgreen}\checkmark} & {\color{red}$\times$} & {\color{red}$\times$} & {\color{darkspringgreen}\checkmark} & - \\
ESP-SoC~\cite{esp_chip} & {\color{darkspringgreen}\checkmark} & {\color{red}$\times$} & {\color{red}$\times$} & Limited & 351 \\
Celerity~\cite{celerity} & {\color{darkspringgreen}\checkmark} & {\color{red}$\times$} & {\color{red}$\times$} & Limited & 80\\
FlexNoC~\cite{flexnoc} & {\color{red}$\times$} & {\color{red}$\times$} & {\color{red}$\times$} & - & -\\
Constellation~\cite{constel} & {\color{darkspringgreen}\checkmark} & {\color{red}$\times$} & {\color{red}$\times$} & {\color{darkspringgreen}\checkmark} & -\\
Andreas et al.~\cite{andy_paper} & {\color{darkspringgreen}\checkmark} & {\color{darkspringgreen}\checkmark} & {\color{darkspringgreen}\checkmark} & {\color{darkspringgreen}\checkmark} & 2146 \\
PATRONoC & {\color{darkspringgreen}\checkmark} & {\color{darkspringgreen}\checkmark} & {\color{darkspringgreen}\checkmark} & {\color{darkspringgreen}\checkmark} & 2700 \\
\bottomrule
\multicolumn{4}{l}{$^{\mathrm{*}}$Normalized to 1~GHz for fair
comparison.}
\end{tabular}
\label{tab:sota}
\end{center}
\vspace{-1em}
\end{table}

\section{Related Work}
\label{sec:relw}
NoCs are an active area of research, and much effort has gone into optimizing topologies, routing algorithms, flow control schemes, and the microarchitecture of routers~\cite{noc_book1,noc_book2,noc_book3}. Multi-core (CPU) architectures have been exploiting these optimizations of NoCs for many decades. However, NoCs for multi-accelerator DNN platforms are still in nascent stage. Table~\ref{tab:sota} provides a brief overview of state-of-the-art NoCs used in multi-core DNN platforms compared to PATRONoC. PATRONoC is the only design that provides open-source AXI-compliant homogeneous burst-based configurable NoC for multi-core DNN platforms. Moreover, PATRONoC outperforms most of the designs in terms of throughput, with the exception of~\cite{cmesh}, which uses a bigger \numproduct{8x8} concentrated mesh (CMesh) topology with primarily local access patterns. Moreover, its results are taken from the gem5 simulator~\cite{gem5}, and the RTL of the design is not openly available. 
Using a CMesh topology for PATRONoC would similarly improve its performance. 

OpenSoC Fabric~\cite{opensoc} is among the few open-source NoCs with a custom non-coherent NoC protocol. It provides a socket to plug AXI-Lite-based endpoints. Unfortunately, AXI-Lite does not support bursts needed by high-performance systems. The ESP framework~\cite{esp,esp_chip} also provides an open-source implementation of its multi-plane NoC, supporting coherent and non-coherent endpoints. The NoC is a 2D mesh topology and uses a custom packet-based protocol. We used ESP-NoC as a baseline for comparison with PATRONoC. Section~\ref{sec:impl} shows that PATRONoC is more area efficient and provides higher bandwidth owing to its homogeneous network. BaseJump Manycore is an open-source non-coherent NoC based on a 2D mesh used in the Celerity chip~\cite{celerity}. Those NoCs are generally limited to meshes and use classical packet-based NoC  protocols, which lead to high area overhead and low bandwidth. In comparison, PATRONoC can be used to design any topology, while providing a highly parameterizable NoC.

\section{Conclusion}
\label{sec:conc}

This work presented the first homogeneous AXI-compliant network-on-chip architecture, building a complete open-source infrastructure for generating various NoC topologies. Using the benefits of a burst-based AXI protocol, PATRONoC targets the emerging field of multi-core DNN platforms requiring high-bandwidth burst-based traffic. The NoC provides high-performance gain compared to state-of-the-art NoCs by using its burst capability and achieves up to a maximum of \qty{310}{\gibi\byte\per\second} aggregated throughput on DNN workloads. The work provides insight into the exploration of different design parameters which affect the performance and complexity of the NoC. It also enables future work to explore different NoC topologies which might be suited for emerging DNN platforms.



\bibliographystyle{IEEEtran}
\bibliography{bstctl, refs}

\clearpage
\newpage

\end{document}